\documentclass[prl,reprint,superscriptaddress]{revtex4-1}

\usepackage{graphicx}
\usepackage{hyperref}
\usepackage{color}
\usepackage[table]{xcolor}
\usepackage{boldline,multirow}
\usepackage{booktabs}
\usepackage{pifont}
\usepackage{mathrsfs}
\usepackage[utf8]{inputenc}
\usepackage{url}

\usepackage[left=2cm, right=2cm, top=2cm, bottom=2cm]{geometry}
\usepackage{xcolor}
\begin{document}

\title{Self-Interacting Dark Matter Subhalos in the Milky Way's Tides}

\author{Omid Sameie}
\email{osame001@ucr.edu}
\affiliation{Department of Physics and Astronomy, University of California, Riverside, CA 92521 USA}
\affiliation{Department of Astronomy, The University of Texas at Austin, 2515 Speedway, Stop C1400, Austin, TX 78712 USA}
\author{Hai-Bo Yu}
\email{haiboyu@ucr.edu}
\affiliation{Department of Physics and Astronomy, University of California, Riverside, CA 92521 USA}
\author{Laura V. Sales}
\affiliation{Department of Physics and Astronomy, University of California, Riverside, CA 92521 USA}
\author{Mark Vogelsberger}
\affiliation{ Department of Physics, Kavli Institute for Astrophysics and Space Research, Massachusetts Institute of Technology, Cambridge, MA 02139, USA\vspace{1pt}}
\author{Jes\'us Zavala}
\affiliation{Center for Astrophysics and Cosmology, Science Institute, University of Iceland, Dunhagi 5, 107 Reykjavik, Iceland
}
\begin{abstract}

We study evolution of self-interacting dark matter (SIDM) subhalos in the Milky Way (MW) tidal field. The interaction between the subhalos and the MW's tides lead to more diverse dark matter distributions in the inner region, compared to their cold dark matter counterparts. We test this scenario with two MW satellite galaxies, Draco and Fornax, opposite extremes in the inner dark matter content, and find that they can be accommodated within the SIDM model proposed to explain the diverse rotation curves of spiral galaxies in the field.

\end{abstract}

\maketitle

\noindent{\textbf{Introduction.}} Dark matter makes up $85\%$ of the mass in the universe, but its nature remains largely unknown. Over the past decades, most studies have focused on the cold dark matter (CDM) model, where the dark matter is composed of collisionless massive particles. CDM has a tremendous success in explaining the large-scale structure of the universe~\cite{Ade:2015xua} and overall features of galaxy formation and evolution~\cite{TG2011, Vogelsberger:2014kha, Vogelsberger2014}. However, it has also long-standing issues in accommodating observations on galactic scales~\cite{Tulin:2017ara,Bullock:2017xww}. Recently, it has been shown that the self-interacting dark matter (SIDM) model~\cite{Spergel:1999mh,Kaplinghat:2015aga} can explain diverse rotation curves of field spiral galaxies~\cite{Kamada:2016euw, Creasey:2016jaq,Ren:2018jpt}, a serious challenge for CDM~\cite{Oman:2015xda,Kaplinghat:2019dhn}. In SIDM, dark matter collisions thermalize the inner halo over the cosmological timescale~\cite{Dave:2000ar,Rocha:2012jg,Peter:2012jh,Vogelsberger:2012ku,vogelsberger2019} and correlate dark matter and baryon distributions~\cite{Kaplinghat:2013xca,Vogelsberger:2014pda,Elbert:2016dbb,Robertson:2017mgj,Sameie:2018chj,Robertson:2018anx, fitts2018}, in accord with observations of spiral galaxies in the field.

SIDM may also provide a solution to puzzles associated with dwarf spheroidal galaxies (dSphs) in the Milky Way (MW), e.g., the most massive subhalos predicted in CDM are too massive to host the bright MW dSphs~\cite{BoylanKolchin:2011de,BoylanKolchin:2011dk}. Simulations show that dark matter self-interactions can lower the central density of the subhalos and alleviate the tension~\cite{Vogelsberger:2012ku,Zavala:2012us,Vogelsberger:2015gpr}. Despite the success, a detailed analysis indicates the preferred dark matter self-scattering cross section per mass, $\sigma/m$, varies within a wide range~\cite{Valli:2017ktb}. This spread reflects diverse dark matter contents of MW dSphs' halos. For example, Draco and Ursa Minor are much denser than Fornax and Sextans~\cite{Valli:2017ktb,Read:2019fxs,Read:2018pft,Kaplinghat:2019yhx}. Unlike spiral galaxies in the field, we expect environmental effects to play a relevant role in shaping MW subhalos. In CDM, the tidal effects can lower central densities for massive subhalos~\cite{Hayashi:2002qv,Penarrubia:2007zx,penarrubia2010,Brooks:2012vi,Wetzel:2016apj} and reduce the number of small ones~\cite{DOnghia:2009xhq,Sawala:2015cdf,Fattahi:2016nld,garrison-kimmel2017,Robles:2019mfq}.

Cosmological simulations show that SIDM subhalos have a larger spread in the inner dark matter content for $\sigma/m=10~{\rm cm^2/g}$~\cite{Vogelsberger:2012ku}, compared to the CDM case. In particular, with such a large cross section, subhalos could experience SIDM core collapse, resulting in high central densities. More recently, it has been suggested that Draco's host could be in the core-collapse phase for $\sigma/m\gtrsim5~{\rm cm^2/g}$~\cite{Nishikawa:2019lsc}, as it has a small pericenter distance to the MW estimated from Gaia data~\cite{Fritz:2018aap} and the tidal effect could trigger the collapse. Intriguingly, for the bright MW dSphs, there is an anti-correlation relation between their central densities and pericenters~\cite{Kaplinghat:2019yhx}, which seems to support this scenario. 

In this {\it Letter}, we explore tidal evolution of SIDM subhalos in the MW's tides, using N-body simulations implemented with realistic MW potentials. The SIDM thermalization coupled with the MW tidal field can lead to diverse dark matter distributions in subhalos. And they can be in either core-collapse or -expansion phases, depending on the cross section, pericenter and initial halo concentration. We explicitly show the mechanism and critical conditions leading to core collapse in the tidal field and study mass loss for both SIDM and CDM subhalos. We further demonstrate that Draco and Fornax, two extremes in the dark matter content among the bright MW dSphs, can be explained in the SIDM model with $\sigma/m=3~{\rm cm^2/g}$, the value used to fit the diverse rotation curves of spiral galaxies in the field~\cite{Kamada:2016euw,Creasey:2016jaq,Ren:2018jpt}.

\begin{figure*}[t!]
\centering{
\includegraphics[scale = 0.25]{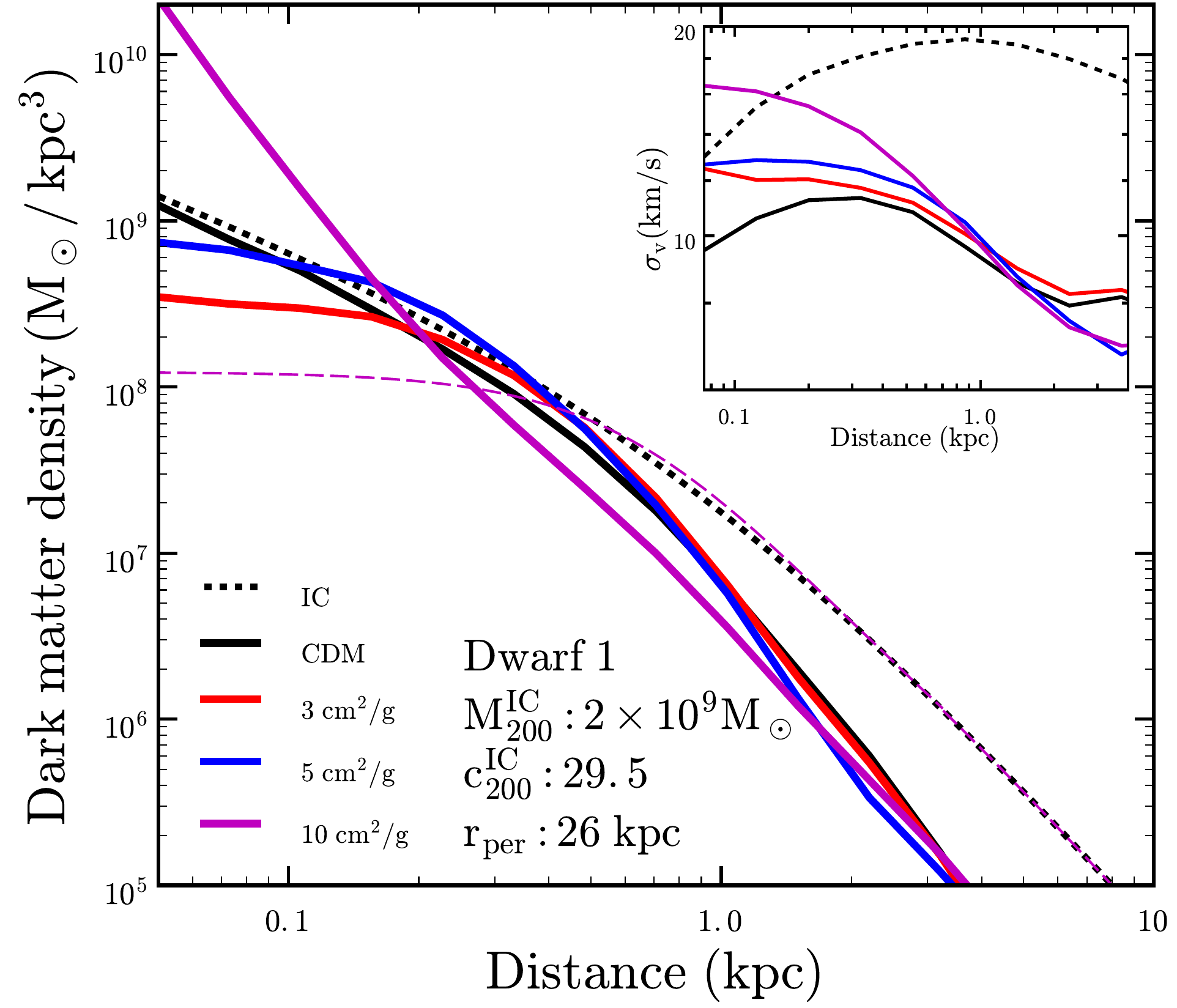}
\includegraphics[scale = 0.25]{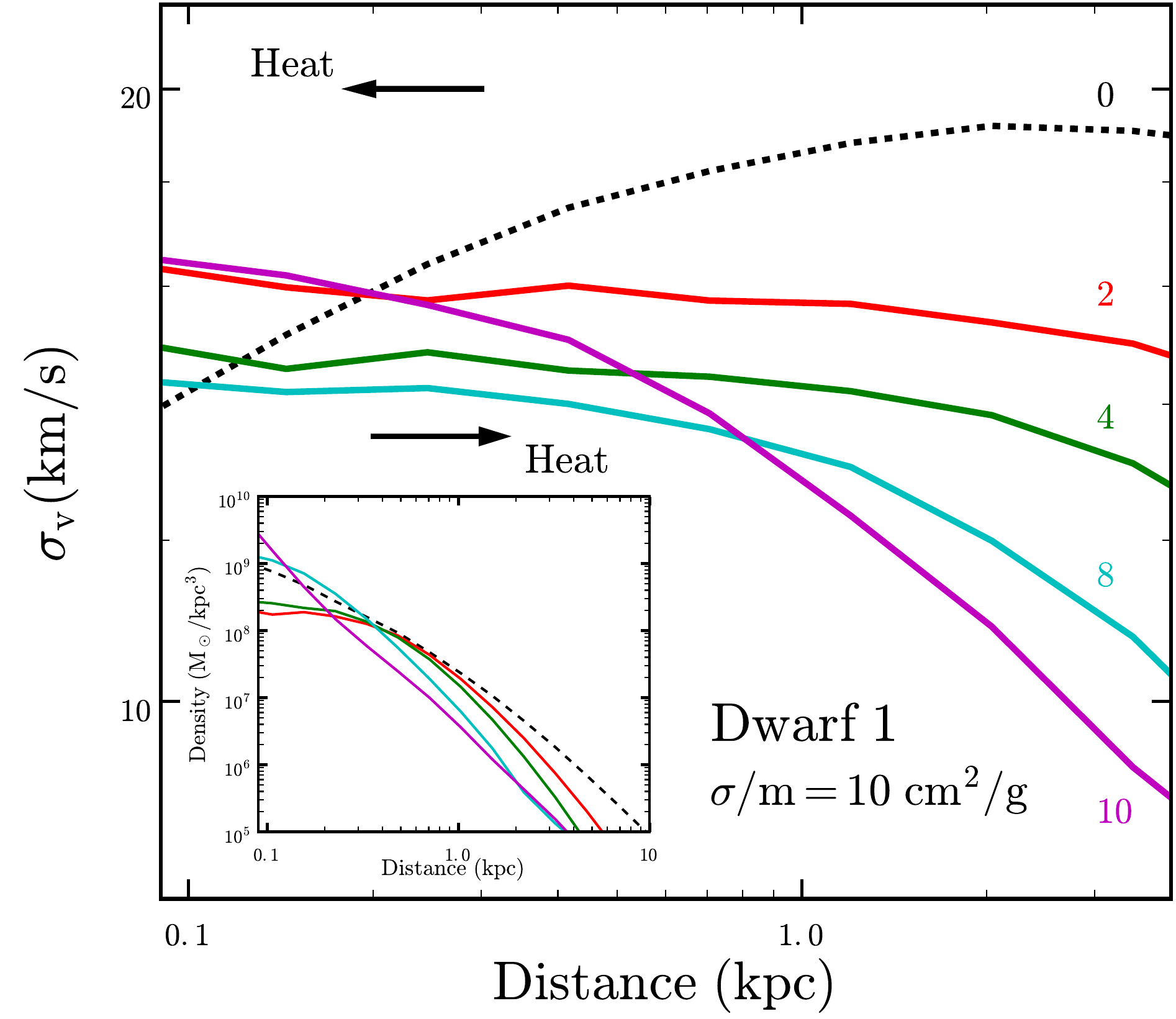}
}
\caption{Left: Dark matter density and velocity-dispersion (inset) profiles at $t=10~{\rm Gyr}$ for Dwarf 1. The magenta dashed curve denotes the density profile assuming Dwarf 1 is an isolated halo with $\sigma/m=10~{\rm cm^2/g}$. Right: dark matter velocity-dispersion and density (inset) profiles at different evolution times, $t=0,~2,~4,~8$ and $10~{\rm Gyr}$ for Dwarf 1 with $\sigma/m=10~{\rm cm^2/g}$. The symbol ``$\leftarrow$" (``$\rightarrow$") denotes the heat-flow direction in the SIDM core-expansion (-collapse) phase.}
\label{fig:fig1}
\end{figure*}

\noindent{\textbf{Simulation setup.}} We carry out N-body simulations using the code {\small\sc AREPO}~\cite{Springel:2000yr} with a module developed in \cite{Vogelsberger:2012ku} for modeling dark matter self-interactions, and use {\small\sc SUBFIND}~\cite{Springel:2000yr} to follow the evolutionary track of the subhalo. Following~\cite{Sameie:2018chj}, we model the baryon and dark matter distributions of the MW with static potentials, while treating dwarf subhalos with the N-body code. This is a good approximation, as we have checked the dynamical friction effect is negligible.

Our MW model includes both disk and bulge components, as they can play a significant role in tidal evolution of the subhalos~\cite{Robles:2019mfq}. For the stellar disk, we use the Miyamoto-Nagai potential $\Phi_{\rm MN}=-GM_{\rm d}/\sqrt{R^2+(R_{\rm d}+\sqrt{z^{2}_{\rm d}+z^2})^2}$~\cite{mn1975}, where $M_{\rm d}=6.98\times10^{10}~{\rm M_\odot}$ is the disk mass, $R_{\rm d}=3.38~{\rm kpc}$ is the disk scale length, and $z_{\rm d}=0.3~{\rm kpc}$ is the disk scale height. We include a Hernquist bulge potential $\Phi_{\rm H}=-GM_{\rm H}/(r+r_{\rm H})$ \cite{hernquist1990}, where $M_{\rm H}=1.05\times10^{10}~{\rm M_{\odot}}$ is the mass and $r_{\rm H}=0.46~{\rm kpc}$. We model the main halo using an NFW profile~\cite{Navarro:1995iw} with the maximum circular velocity $V_{\rm max}=200.5~{\rm km/s}$ and the associated radius $r_{\rm max}=43.4~{\rm kpc}$, and the corresponding halo mass is $M_{200}=1.4\times10^{12}~{\rm M_\odot}$. With these parameters, we can reproduce the MW mass model presented in~\cite{McMillan:2011wd}. In principle, one should also include the self-scattering effect for the main halo. However, for a MW-like galaxy, where the baryons dominate the central regions, an SIDM halo profile can be similar to an NFW one, because SIDM thermalization with the baryonic potential increases the central dark matter density~\cite{Kaplinghat:2013xca,Elbert:2016dbb,Creasey:2016jaq,Sameie:2018chj,Robles:2019mfq}. We have checked that the NFW halo we take here is a good approximation to the SIDM MW halo constructed in~\cite{Sameie:2018chj}. Note the host potential does not evolve with time.

We use an NFW profile to model the initial dark matter distribution in subhalos. To highlight general features of the tidal evolution of SIDM subhalos, in particular critical conditions for the core collapse, we first choose simplified orbital parameters and take the following two sets of initial conditions. Dwarf 1: the halo mass $M_{200}=2\times10^{9}~{\rm M_\odot}$ and concentration $c_{200}=29.5$, or $V_{\rm max}=28.8~{\rm km/s}$ and $R_{\rm max}=1.9~{\rm kpc}$; Dwarf 2:  $M_{200}=2\times10^9~{\rm M_\odot}$ and $c_{200}=22.9$, or $V_{\rm max}=26.7~{\rm km/s}$ and $R_{\rm max}=2.5~{\rm kpc}$. Note Dwarf 2 has the same initial halo mass as Dwarf 1, but its concentration is slightly lower. We use the code {\small\sc S{\footnotesize PHER}IC}~\cite{GarrisonKimmel:2013aq} to generate initial conditions for the subhalos. For both sets of halo parameters, we simulate CDM and SIDM cases with $\sigma/m=3~{\rm cm^2/g}$, $5~{\rm cm^2/g}$ and $10~{\rm cm^2/g}$, and fix the pericenter as $r_{\rm per}=26.5~{\rm kpc}$, motivated by Draco's~\cite{Fritz:2018aap,2018A&A...616A..12G}. We place the initial subhalo at a distance of $230~{\rm kpc}$ from the center of the main halo at $t=0$, and confine the orbit in the plane of the stellar disk. The total number of simulated dark matter particles is $2\times10^6$ in each of our simulated subhalos.

\begin{figure*}[t!]
\centering{
\includegraphics[scale = 0.25]{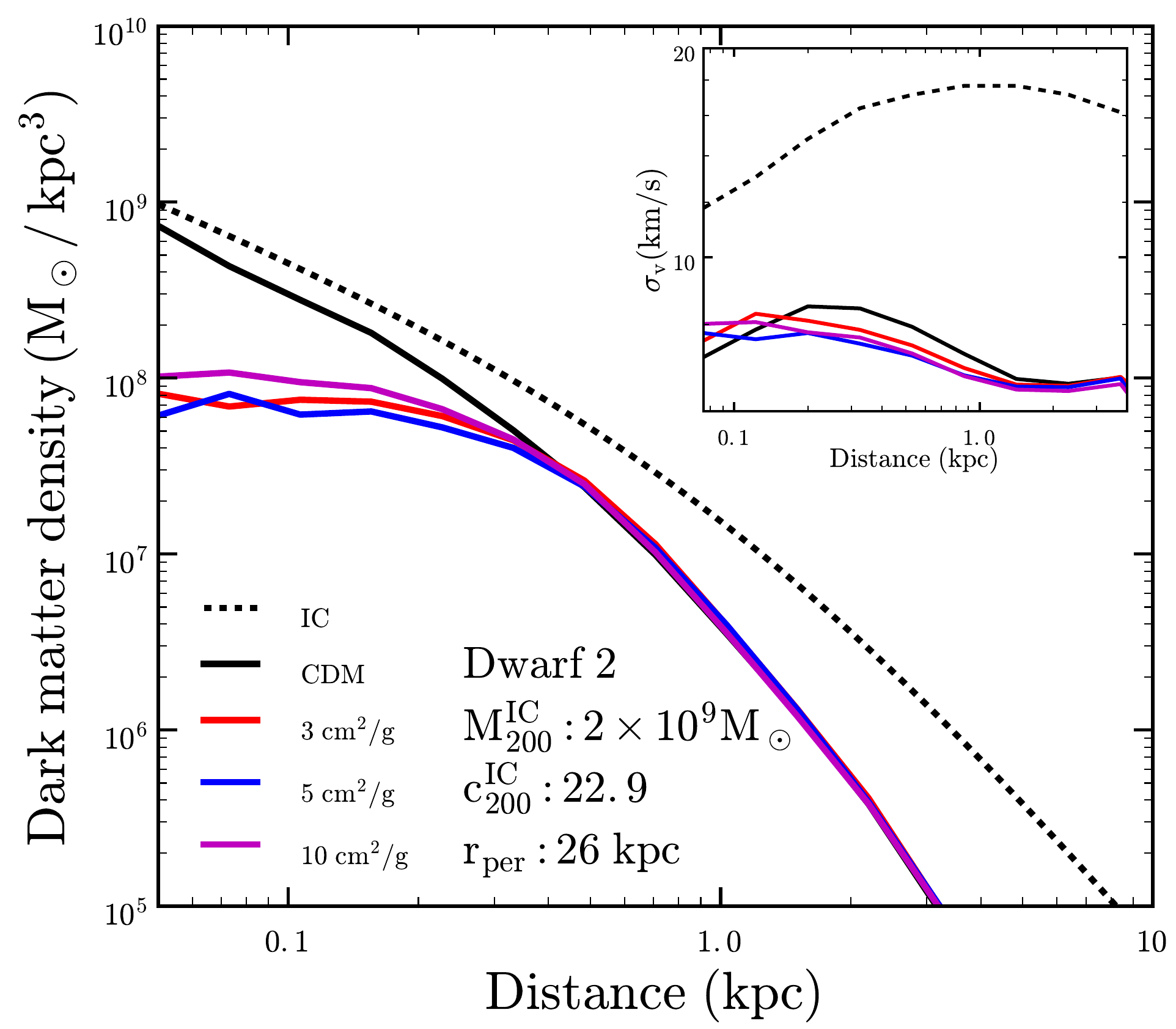}
\includegraphics[scale = 0.25]{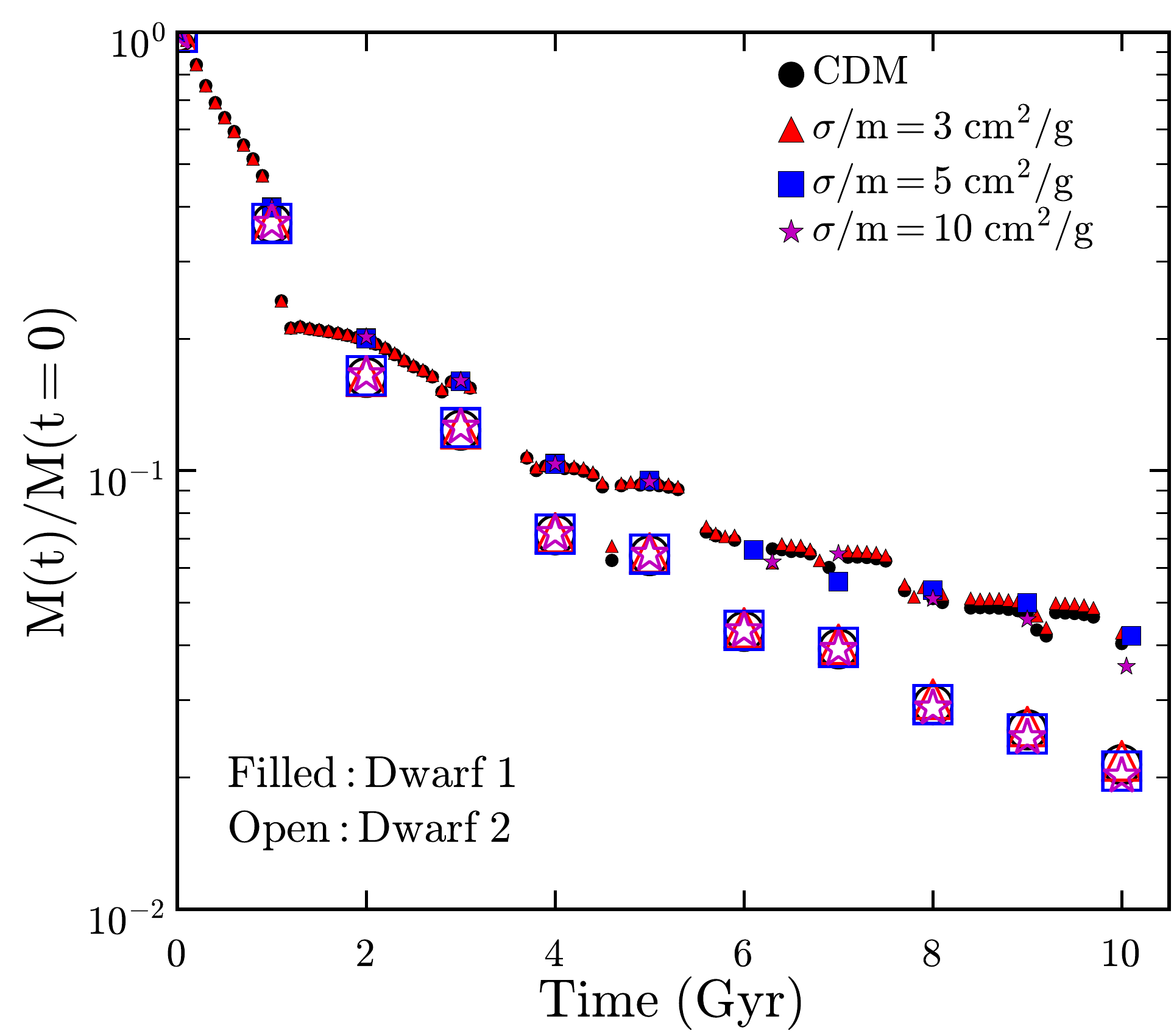}
}
\vspace{0mm}
\caption{Left: Dark matter density and velocity-dispersion (inset) profiles for Dwarf 2. It has the same initial mass as Dwarf 1, but slightly lower concentration. Right: Time evolution of the bound mass, normalized to the initial halo mass. }
\label{fig:fig2}
\vspace{0mm}
\end{figure*}

We then perform two additional simulations to closely model Draco and Fornax in the MW with $\sigma/m=3~{\rm cm^2/g}$. For the Draco subhalo, we take the pericenter $r_{\rm per}=26~{\rm kpc}$, apocenter $r_{\rm apo}=90~{\rm kpc}$, the inclination angle of the orbital plane $\theta=110^{\rm o}$, and the evolution time $t_{\rm evo}=9~{\rm Gyr}$. For the Fornax subhalo, we have $r_{\rm per}=46~{\rm kpc}$, $r_{\rm apo}=150~{\rm kpc}$, $\theta=70^{\rm o}$ and $t_{\rm evo}=6.5~{\rm Gyr}$. These orbital parameters are in a good agreement with measurements using the Gaia DR2 data~\cite{Fritz:2018aap,2018A&A...616A..12G,fillingham2019characterizing,Rocha:2011aa}.  The halo parameters are $M_{200}=4\times10^{9}~{\rm M_\odot}$ and $c_{200}=28$ for Draco; $M_{200}=1.5\times10^{10}~{\rm M_\odot}$ and $c_{200}=12$ for Fornax. The masses are consistent with estimates from abundance matching as in~\cite{Read:2019fxs}. We also include live stellar particles in the simulations and use a Plummer profile to model the initial stellar distribution. The stellar mass is $M_*=2.9\times10^5~{\rm M_\odot}$ and the half-light radius $R_{1/2}=0.22~{\rm kpc}$ for Draco; $M_*=4.3\times10^{7}~{\rm M_\odot}$ and $R_{1/2}=0.71~{\rm kpc}$ for Fornax~\cite{McConnachie:2012vd,deBoer:2012py}. The total number of simulated dark matter particles is the same as before and the numbers of stellar particles are $10^5$ and $5\times10^5$ for Draco and Fornax analogs, respectively.

\noindent{\textbf{Profiles after tidal evolution}} Fig.~\ref{fig:fig1} (left) shows the density and velocity-dispersion profiles (solid) at $t=10~{\rm Gyr}$ for Dwarf 1. In all cases, the MW's tides significantly strip away halo masses and lower densities in the outer regions. All of them have similar density profiles for $r\gtrsim1~{\rm kpc}$, but their central densities are different. For CDM, the inner profile is resilient to tidal stripping and remains cuspy as the initial one (dashed), consistent with earlier findings~\cite{Hayashi:2002qv,Penarrubia:2007zx,penarrubia2010}. While for SIDM, the central density increases with the cross section, opposite to the trend found in field halos~\cite{Elbert:2014bma,Vogelsberger:2015gpr}. In fact, all the SIDM halos are in the core-collapse phase after $10~{\rm Gyr}$'s tidal evolution, as their velocity dispersions are larger than the CDM counterpart and profiles have negative gradients in the inner regions, $r\lesssim1~{\rm kpc}$, an indication of SIDM core collapse~\cite{Balberg:2002ue,Elbert:2014bma,Essig:2018pzq}. For a field halo with the same halo parameters as Dwarf 1,  we use the analytical formula in~\cite{Essig:2018pzq} to estimate the core-collapse time to be $t_{\rm c}\sim16~{\rm Gyr}$ for $\sigma/m=10~{\rm cm^2/g}$, while at $t=10~{\rm Gyr}$ its central density is $1.2\times10^{8}~{\rm M_\odot/kpc^3}$ (magenta dashed), based on the method in~\cite{Kaplinghat:2015aga}. The collapse time becomes even longer for a smaller cross section, as $t_{\rm c}\propto1/(\sigma/m)$. Thus, the MW tides can accelerate the onset of core collapse for SIDM subhalos~\cite{Nishikawa:2019lsc}. We have also checked that for Dwarf 1 with $\sigma/m=3~{\rm cm^2/g}$, the inner density is reduced by $20\%$ if we allow for pre-evolution for $3~{\rm Gyr}$ outside of the main halo and then another $10~{\rm Gyr}$'s tidal evolution.

\noindent{\textbf{Core collapse.}} To appreciate dynamics triggering SIDM core collapse in the tidal field, we take a close look at the evolution history of Dwarf 1 with $\sigma/m=10~{\rm cm^2/g}$, the most extreme case in our study. Fig.~\ref{fig:fig1} (right) shows the velocity-dispersion profiles at different times during the evolution. Overall, the $\sigma_{\rm v}$ value at large radii, say $r=1~{\rm kpc}$, decreases gradually, due to tidal stripping from the MW. Initially, inner $\sigma_{\rm v}$ has a positive gradient in the radius as predicted by the NFW profile. At early stages of the evolution, dark matter self-interactions lead to heat transfer from the outer to inner regions, denoted by the ``$\leftarrow$" symbol, the core size increases and the central density decreases, similar to the case of a field SIDM halo. At the same time, the maximal value of $\sigma_{\rm v}$, the height of the heat reservoir of the dwarf halo, decreases over time due to the mass loss of the dwarf halo in the MW tidal field. Thus, a negative gradient, a necessary condition for the onset of SIDM core collapse, can be more easily satisfied for a subhalo than a field halo. For the example we consider, the transition occurs around $4~{\rm Gyr}$, at a time when the inner dispersion profile is almost flat. Then, the heat flow reverses its direction towards the outer region (``$\rightarrow$"). As a self-gravitating system, the inner halo has negative heat capacity, the more heat is extracted by dark matter collisions, the further it collapses to convert its gravitational energy to kinetic energy~\cite{binney2008}. Thus, both the inner dispersion and density increases at late stages, $t=8\textup{--}10~{\rm Gyr}$. We have also simulated Dwarf 1 with $\sigma/m=10~{\rm cm^2/g}$, but a larger pericenter, $46~{\rm kpc}$, and found the core-collapse transition occurs around $8~{\rm Gyr}$.

\noindent{\textbf{Halo concentration.}} Dwarf 1's initial halo concentration is on the higher end of the distribution predicted in cosmological simulations~\cite{pilipenko2017}. Using the halo concentration-mass relation for field halos at redshift $z=0$ derived in~\cite{Dutton:2014xda}, we find Dwarf 1's $c_{200}$ is $2.5\sigma$ higher than the median value. To  investigate the importance of the concentration in setting the core-collapse timescale, we simulate Dwarf 2 that has the same initial mass as {Dwarf 1} but slightly lower concentration $c_{200}=22.9$. It is $1.5\sigma$ higher than the median in~\cite{Dutton:2014xda}. 

Fig.~\ref{fig:fig2} (left) shows dark matter density and velocity-dispersion profiles for Dwarf 2 after $10~{\rm Gyr}$. We see that all SIDM cases have similar shallow density profiles with the central density $\sim10^{8}~{\rm M_\odot/kpc^3}$. There is no clear evidence of core collapse in Dwarf 2, even though $c_{200}$ is only reduced by $20\%$, compared to Dwarf 1. The result can be understood qualitatively using the scaling relation based on a semi-analytical model, $t_c\propto (\sigma/m)^{-1}M^{-1/3}_{200}c^{-7/2}_{200}$~\cite{Essig:2018pzq}. Dwarf 2's $t_c$ is a factor of $2$ longer than Dwarf 1's due to the small difference in $c_{200}$. Thus, we have demonstrated the critical condition for SIDM core collapse, i.e., a subhalo must have a high concentration. This has important implications for understanding Draco's high dark matter content in SIDM, as we will show later. 

\noindent{\textbf{Mass loss.}} The interplay between dark matter self-interactions and the MW's tides can lead to diverse inner density profiles. However, overall tidal evolution histories for the cases we consider are remarkably similar. Fig.~\ref{fig:fig2} (right) shows the ratio of the total mass of bound particles to the initial halo mass vs. time. In all cases, the halo loses $80\%$ of its initial mass within the first $2~{\rm Gyr}$. Moreover, for a given initial halo and its pericenter, the mass loss rate is almost independent of the self-scattering cross section for the cases we study. We also find that the mass-loss rate is sensitive to the halo concentration. Dwarf 1 and Dwarf 2 have the same pericenter and initial mass, but the former is a factor of $2$ more massive than the latter after $t=10~{\rm Gyr}$'s tidal evolution as Dwarf 1 has a higher initial $c_{200}$ value. These results reflect the fact that a subhalo with high concentration is more resilient to tidal stripping.

\begin{figure}[t!]
\centering{
\includegraphics[scale = 0.25]{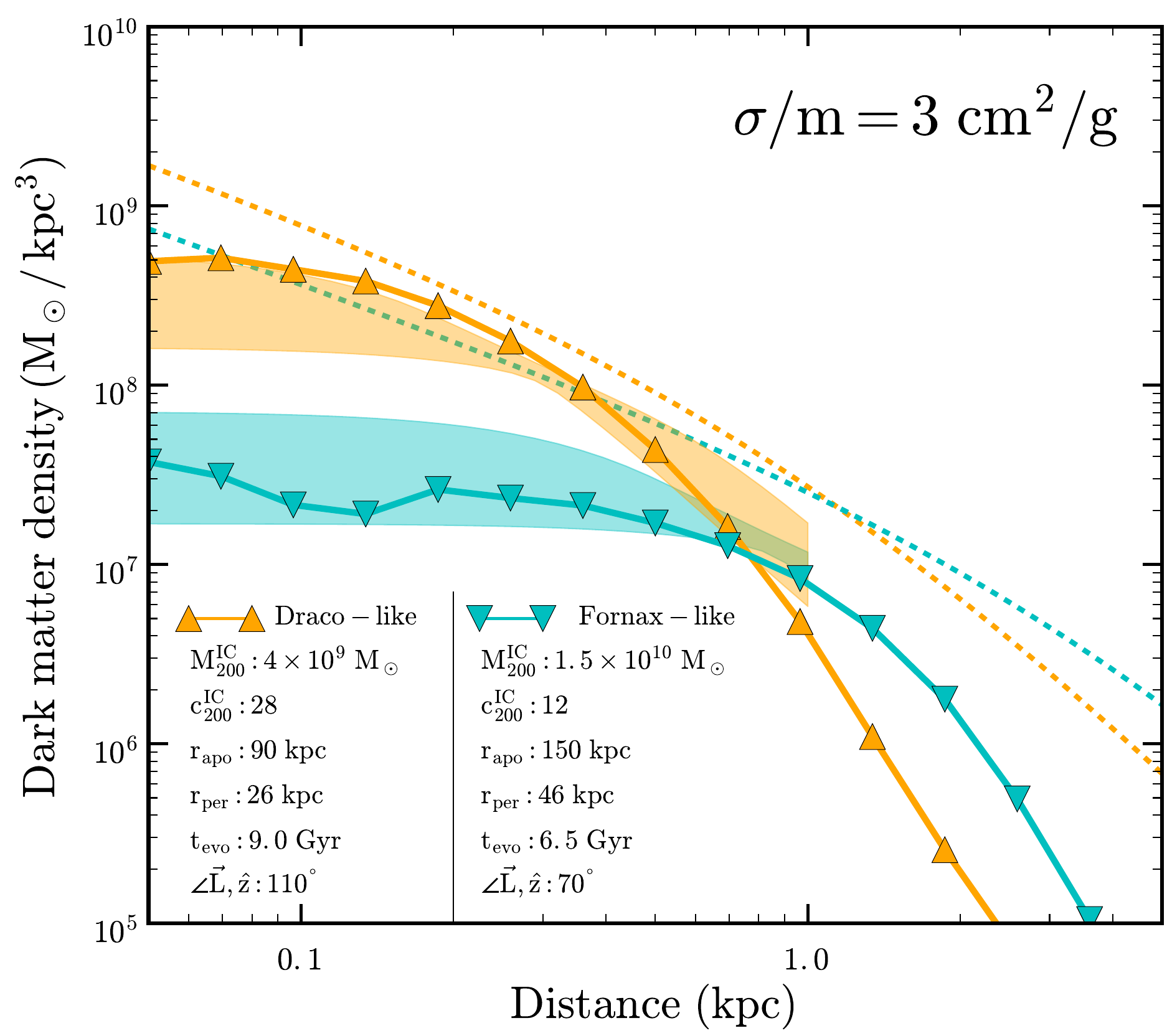}
}
\vspace{0mm}
\caption{Dark matter density profiles of Draco (orange) and Fornax (cyan) analogs in our simulations, where the orbital parameters are consistent with measurements using the Gaia DR2 data~\cite{Fritz:2018aap,2018A&A...616A..12G,fillingham2019characterizing}. The dashed curves denote initial dark matter density profiles. The shaded bands show cored isothermal density profiles from the fits to stellar kinematics of Draco and Fornax at $95\%$ CL~\cite{Kaplinghat:2019yhx}. }
\label{fig:fig3}
\vspace{0mm}
\end{figure}

\noindent{\textbf{A case for Draco and Fornax.}} Fig.~\ref{fig:fig3} shows the density profiles for Draco (orange) and Fornax (magenta) analogs from our simulations, where we take $\sigma/m=3~{\rm cm^2/g}$ for the SIDM runs. The simulated dark matter density profiles agree well with those inferred from the stellar kinematics of Draco and Fornax~\cite{Kaplinghat:2019yhx}. The Draco subhalo has a high concentration, $2.5\sigma$ above the median~\cite{Dutton:2014xda}, and it experiences core collapse as Dwarf 1, resulting in a high central density. We have further checked that even in case of CDM Draco's host has a similar $c_{200}$ value to fit the data. What we have shown is that with the same high concentration SIDM can also produce a high central density due to core collapse triggered by tidal stripping. The Fornax subhalo has a higher initial mass but lower concentration, close to the median. It is still in the core-expansion phase after tidal evolution and dark matter self-interactions lead to a shallow density profile. The total stellar masses after the evolution are $1.9\times10^5~{\rm M_\odot}$ and $2\times10^7~{\rm M_\odot}$ for the Draco and Fornax analogs, respectively, overall consistent with the observations.

If the Draco-like subhalo is in the field, the central density will be $1.1\times10^8~{\rm M_\odot/kpc^3}$, too low to be consistent with the observations. This explains why the earlier analyses~\cite{Valli:2017ktb,Read:2018pft}, where they did not model the core collapse, found $\sigma/m\lesssim0.3~{\rm cm^2/g}$ for Draco. We also note cosmological simulations of a MW-like system in~\cite{Robles:2019mfq} do not show the evidence of core collapse in subhalos. This is because the cross section of $1~{\rm cm^2/g}$ is too low to induce core collapse in even their most concentrated subhalos. It's worth emphasizing that the core collapse of SIDM subhalos has been observed in other cosmological simulations~\cite{Vogelsberger:2012ku}, where one of $15$ top massive subhalos experiences the collapse for $\sigma/m=10~{\rm cm^2/g}$.

We have demonstrated that the interplay between SIDM thermalization and tidal stripping can lead to diverse central densities for subhalos in accord with observations. Our analyses indicate Draco's host halo must have a high concentration. As inferred from observations, it has the most dense inner halo among the nine bright satellite galaxies of the MW~\cite{Kaplinghat:2019yhx,Read:2018fxs}. Further taking into account the population of the ultra-faint satellites, Draco stands out as an overdense subhalo in both CDM and SIDM scenarios. To fully determine the likelihood of accretion of such highly-concentrated halos, cosmological simulations with a statistically significant number of hosts will be necessary. Our results in this work provide useful constraints on the infall properties of the satellite galaxies.

\noindent{\textbf{Conclusions.}} We have shown that the interaction between the SIDM subhalos and the MW's tides can lead to diverse dark matter density profiles. In particular, our simulations show the SIDM core-collapse condition is extremely sensitive to the initial halo concentration. We demonstrated that the SIDM model with a fixed cross section, proposed for field galaxies, can accommodate the MW dSphs Draco and Fornax as well, although their dark matter contents differ significantly. For the cases we studied, the overall mass loss rates are almost identical for SIDM and CDM subhalos. In the future, we could explore the stellar distribution of MW dSphs and its correlation with the core size and the pericenter, including the ultra-faint satellites; see~\cite{Kaplinghat:2019yhx,Kahlhoefer:2019oyt}. It would also be interesting to perform hydrodynamical simulations of a MW-like galaxy and study the tidal effects on SIDM subhalos in the cosmological setup.

\textbf{Acknowledgements}: We thank Manoj Kaplinghat, Matthew Walker and Mauro Valli for useful discussion, and Volker Springel for making {\small \sc Arepo} available for this work. OS acknowledges support from NASA MUREP Institutional Opportunity (MIRO) grant number NNX15AP99A and thanks the Carnegie Observatories for hospitality during the completion of this work. HBY acknowledges support from U.S. Department of Energy under Grant No.~de-sc0008541, UCR Regents' Faculty Development Award, and in part by the U.S. National Science Foundation under Grant No. NSF PHY-1748958 through the ``From Inflation to the Hot Big Bang" KITP program. LVS is grateful for support from the Hellman Fellow Foundation. MV acknowledges support through an MIT RSC award, a Kavli Research Investment Fund, NASA ATP grant  NNX17AG29G, NSF grants AST-1814053 and  AST-1814259. JZ acknowledges support by a Grant of Excellence from the Icelandic Research Fund (grant number 173929-051). Part of the simulations was run on the Extreme Science and Engineering Discovery Environment (XSEDE) and the Comet supercomputer at the San Diego Supercomputer Center, which are funded by the National Science Foundation awards OAC-1548562 and OAC-1341698, respectively.

\bibliographystyle{apsrev4-1}
\bibliography{main_biblio}
\end{document}